\documentclass[prl,twocolumn,showpacs,preprintnumbers,nofootinbib]{revtex4-1}
\usepackage{amsfonts,amsmath,amssymb}
\usepackage{graphicx}
\usepackage{bm}
\usepackage{braket}
\usepackage[caption=false]{subfig}
\usepackage{xcolor}
\usepackage{comment}

\newcommand{ \mysmall}[1]{\scriptscriptstyle \mathrm{#1}} 
\newcommand{\had}{\mathrm{h}}
\newcommand{\lep}{\mathrm{l}}
\newcommand{ \eq}[1]{Eq.~(\ref{eqn:#1})}

\renewcommand{\Im}{\mathrm{Im}}

\begin{document}
\preprint{SI-HEP-2018-38, QFET-2018-25}

\title{Muon-electron scattering at NNLO: the hadronic corrections}

\author{M.~Fael}
\email{fael@physik.uni-siegen.de}
\affiliation{Theoretische Physik I, Universit\"{a}t Siegen, 57068 Siegen, Germany}

\author{M.~Passera} 
\email{passera@pd.infn.it}
\affiliation{Istituto Nazionale Fisica Nucleare, Sezione di Padova, I-35131 Padova, Italy}

\begin{abstract} 
\noindent 
The Standard Model prediction for muon-electron scattering beyond leading order requires the inclusion of QCD contributions which cannot be computed perturbatively. At next-to- and next-to-next-to-leading order, they arise from one- and two-loop diagrams with hadronic vacuum polarization insertions in the photon propagator. We present their evaluation using the dispersive approach with hadronic $e^+e^-$ annihilation data and estimate their uncertainty. We find that these corrections are crucial for the analysis of future high-precision muon-electron scattering data, like those of the recently proposed MUonE experiment at CERN.

~

\end{abstract} 

\date{\today}
\maketitle

\section{Introduction}

The elastic scattering of muons and electrons is one of the most basic processes in particle physics. It has been studied since the late 1930s, when measurements of the collisions of muons in cosmic rays with atomic electrons played a crucial role in the discovery of the muon. In spite of this long history,  few measurements are available. In the 1960s, the $\mu e$ elastic scattering cross section was measured at CERN and Brookhaven using accelerator-produced muons~\cite{Backenstoss:1963,Kirk:1968,Jain:1969wt}, and in the late 1990s the scattering of muons off polarized electrons was used by the SMC collaboration at CERN as a polarimeter for high-energy muon beams~\cite{Adams:1999af}.

Recently, a novel approach has been proposed to determine the leading hadronic contribution to the muon $g$-2, measuring the effective electromagnetic coupling in the spacelike region via scattering data~\cite{Calame:2015fva}. The elastic scattering of high-energy muons on atomic electrons has been identified as an ideal process for this measurement, and a new experiment, MUonE, has been proposed at CERN to measure the differential cross section of $\mu e$ elastic scattering as a function of the spacelike squared momentum transfer~\cite{Abbiendi:2016xup}. In order for this new determination of the leading hadronic corrections to the muon $g$-2 to be competitive, the $\mu e$ differential cross section must be measured with statistical and systematic uncertainties of the order of 10ppm. An analogous precision is therefore required in the theoretical prediction.

Until recently, the Standard Model (SM) prediction of the $\mu e \to \mu e$ process had received little attention. Only the next-to-leading order (NLO) QED corrections to the differential cross section were computed (long time ago) in~\cite{Nikishov:1961,Eriksson:1961,Eriksson:1963,VanNieuwenhuizen:1971yn,DAmbrosio:1984abj,Kukhto:1987uj,Bardin:1997nc} and revisited more recently in~\cite{Kaiser:2010zz}. As a first check, we recalculated these corrections and found agreement with~\cite{Kaiser:2010zz} (see also~\cite{MarcoRocco2017,MarcoVitti2018}). An important step forward was taken very recently by the authors of~\cite{Alacevich:2018vez}, who calculated the full set of NLO QED corrections without any approximation and developed a fully differential Monte Carlo code. They also computed the full set of NLO electroweak corrections.

The QED corrections at next-to-next-to-leading order (NNLO), crucial to interpret the high-precision data of future experiments like MUonE, are not yet known, although some of the two-loop corrections which were computed for Bhabha scattering in QED~\cite{Bern:2000ie,Bonciani:2003te,Bonciani:2003cj} and for $t{\bar t}$ production in QCD~\cite{Bonciani:2008az, Bonciani:2013ywa} can be applied to $\mu e$ scattering as well. A first step towards the calculation of the full NNLO QED corrections to $\mu e$ scattering was taken in~\cite{Mastrolia:2017pfy,Mastrolia:2018sso,DiVita:2018nnh}, where the master integrals for the two-loop planar and non-planar four-point Feynman diagrams were computed. These integrals were calculated setting the electron mass to zero, while retaining full dependence on the muon one. The extraction of the leading electron mass effects from the massless $\mu e$ scattering amplitudes has been recently addressed in~\cite{Engel:2018fsb} (see also~\cite{Penin:2005kf,Mitov:2006xs,Becher:2007cu}).

In this letter we will study the hadronic corrections to $\mu e$ scattering. While at NLO these corrections are simply proportional to the product of the LO QED predictions and the hadronic part of the vacuum polarization, at NNLO their evaluation is complicated by the presence of non-factorizable two-loop diagrams. We will present their calculation using the dispersive approach with hadronic $e^+e^-$ annihilation (timelike) data~\cite{Cabibbo:1961sz}. This approach was used, for example, to calculate the hadronic corrections to muon decay~\cite{vanRitbergen:1998hn,Davydychev:2000ee} and Bhabha scattering~\cite{Actis:2007fs,Kuhn:2008zs,CarloniCalame:2011zq}. We point out that, taking advantage of the hyperspherical integration method, it was recently shown that these non-factorizable diagrams can also be calculated employing the hadronic vacuum polarization in the spacelike region, without using timelike data~\cite{Fael:2018dmz}. We will conclude that the hadronic NNLO corrections to $\mu e$ scattering are important in comparison with the expected accuracy of future high-precision experiments like MUonE, and we will show how to properly include them in their analysis.

\section{Details of the calculation}
The SM prediction for the unpolarized differential cross section of the elastic scattering 
\begin{align}	
	\mu^{\pm} e^- \to \mu^{\pm} e^-
\end{align}
at leading order is 
\begin{equation}
	\frac{d \sigma_0}{dt}  =  4 \pi \alpha^2 \,
	\frac{\left(M^2+m^2\right)^2 - su + t^2/2}{t^2 \lambda \left(s,M^2,m^2 \right)}, 
\label{eqn:sigmaLO}
\end{equation}
where $m$ ($M$) is the electron (muon) mass, $s, t, u$ are the Mandelstam variables satisfying $s+t+u=2m^2+2M^2$, $\alpha$ is the fine-structure constant, and $\lambda(x,y,z)=x^2+y^2+z^2-2xy-2xz-2yz$ is the K{\"a}llen function.

In a fixed-target experiment where the electron is initially at rest, $E_{\mu}$ is the energy of the incoming muons or antimuons, and $E'_e$ is the electron recoil energy, the variables $s$ and $t$ are given by 
\begin{align}
	&s \, = \,  2mE_{\mu}+M^2+m^2, \\
	& t  \, = \, -2m(E'_e-m), \\
	& t_{\rm min}  <  \, t  < 0, \\
	&t_{\rm min} = -\lambda(s,M^2,m^2)/s.
\end{align}	
Both positive and negative muon beams will be available for the MUonE experiment. For $E_{\mu} = 150$~GeV, which is a typical energy available at the M2 beam line in CERN's North Area, $s = 0.164$~GeV$^2$ and $-0.143~{\rm GeV}^2 < t < 0$.

To evaluate the hadronic corrections to $\mu e$ scattering at NLO and NNLO, let us consider the SM vacuum polarization tensor with four-momentum $q$,
\begin{align}
  i \Pi^{\mu\nu}(q) &= i \Pi(q^2) \left(g^{\mu\nu}q^2-q^\mu q^\nu \right) =  i \Pi(q^2) \, q^2 t^{\mu\nu} (q) \notag \\
  &=  \int d^4 x \,\, e^{iqx} 
  \bra{0} T \left\{j^\mu_{\mysmall{em}}(x) j^\nu_{\mysmall{em}}(0) \right\} \ket{0},
\end{align}
where $j^\mu_{\mysmall{em}}(x) = \sum_f Q_f \, \bar{\psi_f}(x) \gamma^\mu \psi_f(x)$ is the electromagnetic current and the sum runs over fermions with charges $Q_f$. The weak interactions will be ignored. The transverse part of the full photon propagator is
\begin{equation}
  \frac{-i t^{\mu\nu}}{q^2 \left[1+\Pi(q^2) \right]}
  = \frac{-i t^{\mu\nu}}{q^2}
  \Bigg[ 1-\Pi(q^2) + \Pi^2(q^2) +  \dots \Bigg],
  \label{eqn:dysonserie}
\end{equation}
where $\Pi(q^2)$ is the renormalized vacuum polarization function satisfying the condition $\Pi(0) = 0$. It receives contributions from the charged leptons (l), the five light quarks $u,d,s,c,b$ with the corresponding hadrons (h), and from the top quark:
\begin{equation}
  \Pi(q^2) = 
  \Pi_\lep(q^2)
  +\Pi_\had(q^2)
  +\Pi_\mathrm{top}(q^2).
\end{equation}
While the purely leptonic contribution $\Pi_\lep(q^2)$, as well as $\Pi_\mathrm{top}(q^2)$, involving only the top quark, can be computed order by order in $\alpha$ and $\alpha_s$~\cite{Steinhauser:1998rq,Sturm:2013uka,Kuhn:1998ze}, the hadronic one cannot be calculated in perturbation theory for any value of $q^2$, because of the non-perturbative nature of the strong interaction at low energy. Yet, the subtracted dispersion relation
\begin{equation}
  \frac{\Pi_\had(q^2)}{q^2} =
  - \frac{1}{\pi}
  \int_{4m_\pi^2}^{\infty}
  \frac{dz}{z} \frac{\Im \Pi_\had(z+i\varepsilon)}{q^2-z+i\varepsilon}
  \label{eqn:dispersionrelation}
\end{equation}
and the optical theorem $\Im \Pi_\had(s) = (\alpha/3) R(s)$, where
\begin{align}
  R(s) = 
  \sigma(e^+e^- \to \text{hadrons}) /\frac{4\pi |\alpha(s)|^2}{3 s}
\end{align}
and
\begin{align}
	\alpha (s) = \frac{\alpha}{1- \Delta \alpha (s)}
\end{align}
is the effective fine-structure constant, allow to express the hadronic vacuum polarization in terms of the measured cross section of the reaction $e^+ e^- \to $ hadrons~\cite{Jegerlehner:2017gek}.

\begin{figure}[htb]
\centering
\subfloat[NLO \label{fig:hadlov2}]{
    \includegraphics[width=0.32\columnwidth]{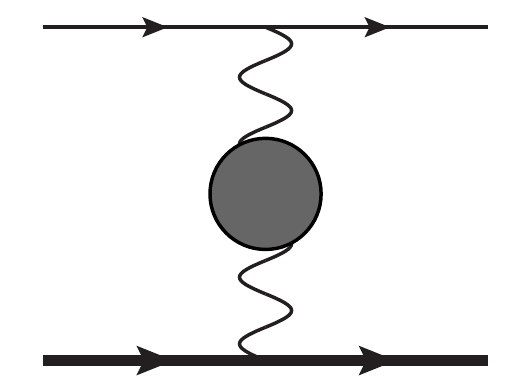}}
    \subfloat[class I\label{fig:classI}]{
    \includegraphics[width=0.32\columnwidth]{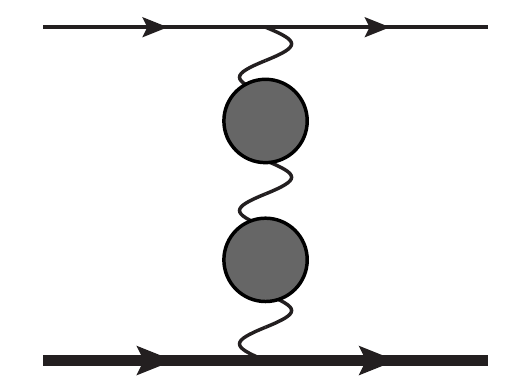}}
  \subfloat[class II\label{fig:classII}]{
    \includegraphics[width=0.32\columnwidth]{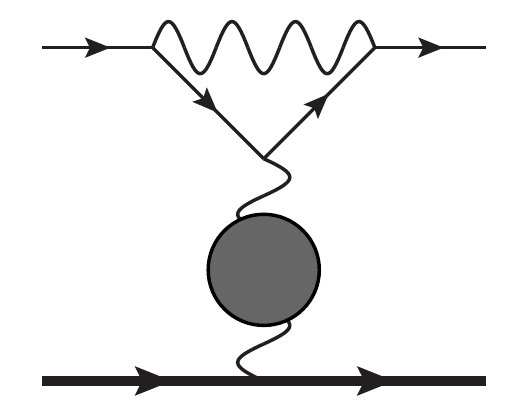}} \\
  \subfloat[class III\label{fig:classIII}]{
    \includegraphics[width=0.32\columnwidth]{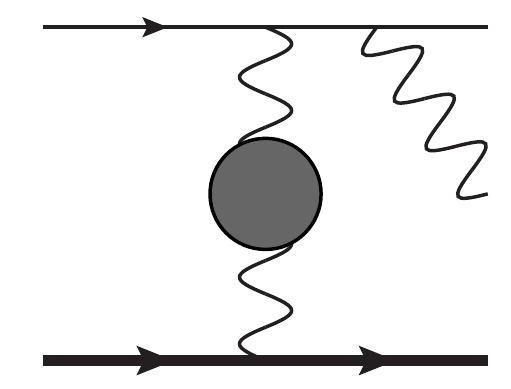}} 
  \subfloat[class IV\label{fig:classIV}]{
  \includegraphics[width=0.32\columnwidth]{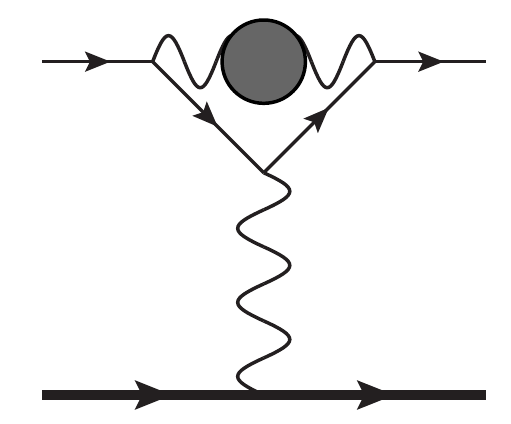}
  \includegraphics[width=0.32\columnwidth]{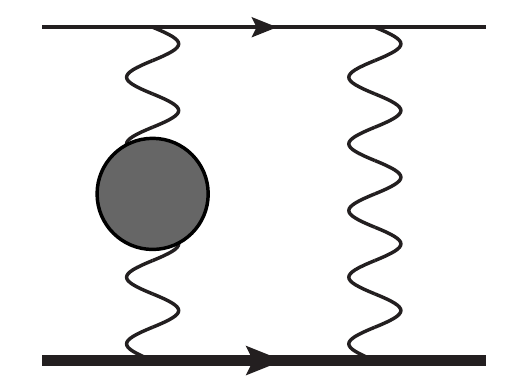}}
  \caption{(a) Diagram contributing to the hadronic correction to $\mu e$ scattering at NLO. (b--e) Examples of diagrams contributing to the four classes of hadronic corrections at NNLO. Electrons, muons and photons are depicted with thin, thick and wavy lines, respectively. The grey blobs indicate hadronic vacuum polarization insertions.}
  \label{fig:fd}
\end{figure}

At NLO, the hadronic correction to the $\mu e$ differential cross section, of order $\alpha^3$, is due to the diagram in Fig.~\ref{fig:hadlov2}. It is the same for positive and negative muon beams:
\begin{equation}
	\frac{d \sigma_\had^{\mathrm{NLO}}}{dt}  =  - 2 \Pi_\had(t) \, \frac{d \sigma_0}{dt}.
\label{eqn:sigmaHNLO}
\end{equation}
As the leading hadronic contribution to the running of $\alpha(t)$ is $\Delta \alpha_\had(t) = -\Pi_\had(t)$, \eq{sigmaHNLO} accounts for the leading hadronic effect of the running of the electromagnetic coupling constant in the spacelike region. The extraction of this quantity from $\mu e$ scattering data is, therefore, the goal of the MUonE experiment. Alternatively, its numerical value can be obtained using the Fortran libraries \texttt{alphaQEDc17}~\cite{Jegerlehner:2001ca,Jegerlehner:2006ju,Jegerlehner:2011mw,Harlander:2002ur} and \texttt{KNT18VP}~\cite{Hagiwara:2003da,Hagiwara:2006jt,Actis:2010gg,Hagiwara:2011af,Keshavarzi:2018mgv,Harlander:2002ur} based on hadronic $e^+ e^-$ annihilation (timelike) data. Let us define the ratio
\begin{equation}
	K_\had^{\mathrm{NLO}}(t) = \frac{d \sigma_\had^{\mathrm{NLO}}}{dt} / \frac{d \sigma_0}{dt} =  - 2 \Pi_\had(t).
\label{eqn:RhoHNLO}
\end{equation}
For $E_{\mu} = 150$~GeV, it reaches the maximum value of $2.1 \times 10^{-3}$ at $t=t_{\rm min}=-0.143~{\rm GeV}^2$. For the same incoming muon energy, the maximum value of the top quark contribution is a tiny $2.0 \times 10^{-9}$.

At NNLO, the hadron-induced corrections to $\mu e$ scattering, of order $\alpha^4$, can be split into four classes of diagrams:
\begin{enumerate}
\item[I.] Class I consists of tree level diagrams with one or two vacuum polarization insertions (Fig.~\ref{fig:classI}). Their contribution to the differential cross section is proportional to $\Pi_\had(t) \left[ \Pi_\had(t) + 2 \Pi_\lep(t) \right]$, the reducible part of the second-order hadronic contribution to the running of $\alpha(t)$. We note that, in Fig.~\ref{fig:hadlov2}, a virtual photon can be emitted and reabsorbed by the hadronic insertion. This irreducible part of the second-order hadronic contribution to the running of $\alpha(t)$ will not be considered as part of class I (although of the same order), because its effect is commonly included in the ratio $R(s)$ as final-state radiation and, therefore, it is already incorporated in the NLO hadronic corrections in \eq{sigmaHNLO}~\cite{Melnikov:2001uw,Passera:2004bj}.
\item[II.] QED one-loop diagrams in combination with one hadronic vacuum polarization insertion in the $t$-channel photon (Fig.~\ref{fig:classII}). Their contribution to the differential cross section is proportional to $\Pi_\had(t)$ and a combination of one-loop QED corrections to $\mu e$ scattering.
\item[III.] Real photon emission diagrams with a vacuum polarization insertion in the $t$-channel photon (Fig.~\ref{fig:classIII}). They contain terms proportional either to $\Pi_\had(t_e)$ or to $\Pi_\had(t_\mu)$, where $t_e$ ($t_\mu$) is the square of the difference of the initial and final electron (muon) momenta. In general, $t_e \neq t_\mu $ because of the presence of the final-state photon. 
\end{enumerate}
All the diagrams in classes I--III are \emph{factorizable}, since each of them can be reduced to the product of a QED amplitude multiplied by the function $\Pi_\had(q^2)$ evaluated at some $q^2$ value fixed by the external kinematics. A fourth class of \emph{non-factorizable} diagrams must also be considered:
\begin{enumerate}
\item[IV.] One-loop QED amplitudes with a hadronic vacuum polarization insertion in the loop. They can be further subdivided into vertex and box corrections (Fig.~\ref{fig:classIV}).
\end{enumerate}
We point out that there are no light-by-light contributions to the $\mu e$ cross section at NNLO (order $\alpha^4$) -- they appear at N${}^{3}$LO (order $\alpha^5$). Moreover, we remind the reader that, at the level of precision addressed in this letter, the analysis of future $\mu e$ scattering data will also require the study of $\mu e$ scattering processes with final states containing hadrons. Final states of Bhabha scattering containing hadrons were studied in~\cite{CarloniCalame:2011zq}.

We calculated the amplitudes in class IV employing the dispersion relation in~\eq{dispersionrelation}. The factor $\Pi_\had(q^2)/q^2$ appearing in the loop -- where $q$ now stands for the loop momentum -- is replaced by the r.h.s.\ of \eq{dispersionrelation}, where $q$ appears only in the denominator of the term $1/(q^2-z)$. Therefore, the dispersion relation effectively replaces the dressed propagator with a massive one, where $z$ plays the role of a fictitious squared photon mass. This allows to interchange the integration order and evaluate, as a first step, the one-loop amplitudes with a ``massive" photon. The results obtained for the $z$-dependent scattering amplitudes are then convoluted with the ratio $R(s)$.

All four classes of diagrams were generated using \texttt{FeynArts}~\cite{Hahn:2000kx} with a modified version of the QED model that contains, besides leptons and photons, a fictitious massive gauge boson (the ``massive" photon arising from the dispersion relation). The amplitudes were calculated and reduced to one-loop tensor integrals with \texttt{Form}~\cite{Kuipers:2012rf} via the \texttt{FormCalc}~\cite{Hahn:1998yk} package, and exported as a Fortran code for the numerical evaluation of the dispersive and phase-space integrals. Two independent parametrizations of the 3-body phase space were employed to cross-check the hard bremsstrahlung cross section. 
For the numerical evaluation of $\Pi_\had(q^2)$ in the spacelike region, appearing in classes I--III, we relied on the native implementation available in the Fortran libraries \texttt{alphaQEDc17} and \texttt{KNT18VP}.
The one-loop tensor coefficients were computed with the library \texttt{Collier}~\cite{Denner:2016kdg}, which features dedicated expansions for the evaluation in numerically unstable regions (small Gram or other kinematical determinants). We particularly benefited from this library when we convoluted the $z$-dependent amplitudes with the $R(s)$ ratio provided by~\texttt{alphaQEDc17} or \texttt{KNT18VP}. Indeed, in performing the dispersive integrations in class IV diagrams, the squared photon ``mass" $z$ appearing inside the loop functions can acquire values which are orders of magnitude larger than the typical energy scale of the scattering process. \texttt{Collier} provides numerically stable results in this treacherous region and allows the numerical integration to converge.
The dispersive integrations were performed with the subroutines in \texttt{QUADPACK}~\cite{piessens1983quadpack}, while for the phase space integration we employed the \texttt{VEGAS} algorithm~\cite{Lepage:1977sw} in the \texttt{Cuba} library~\cite{Hahn:2004fe}.

To check our results, we produced an independent \textsc{Mathematica} implementation using \texttt{FeynCalc}~\cite{Mertig:1990an,Shtabovenko:2016sxi} and \texttt{Package-X}~\cite{Patel:2015tea}. The results obtained by \texttt{FeynCalc} in terms of scalar one-loop functions were then evaluated numerically using analytic expressions provided by \texttt{Package-X}. The use of \textsc{Mathematica}'s arbitrary-precision numbers, with a large number of digits, allowed us to keep track of precision at all steps and avoid instabilities during the numerical dispersive and phase-space integrations. We found perfect agreement between the two implementations.

The lepton masses were kept different from zero throughout the calculation, so that the matrix elements were free of collinear singularities. Ultraviolet singularities were regularized via conventional dimensional regularization and UV-finite results were obtained in the on-shell renormalization scheme. The amplitudes of class II and the boxes of class IV develop IR poles which are cancelled by those arising from the phase space integration of the real emission diagrams of class III. We employed both the FKS subtraction scheme~\cite{Frixione:1995ms,Frederix:2009yq} as well as the traditional QED procedure to assign a vanishingly small mass to the photon to remove the soft singularities and to obtain an IR-finite cross section.

\section{Results}
\begin{figure}[htb]
\centering
    \includegraphics[width=1\columnwidth]{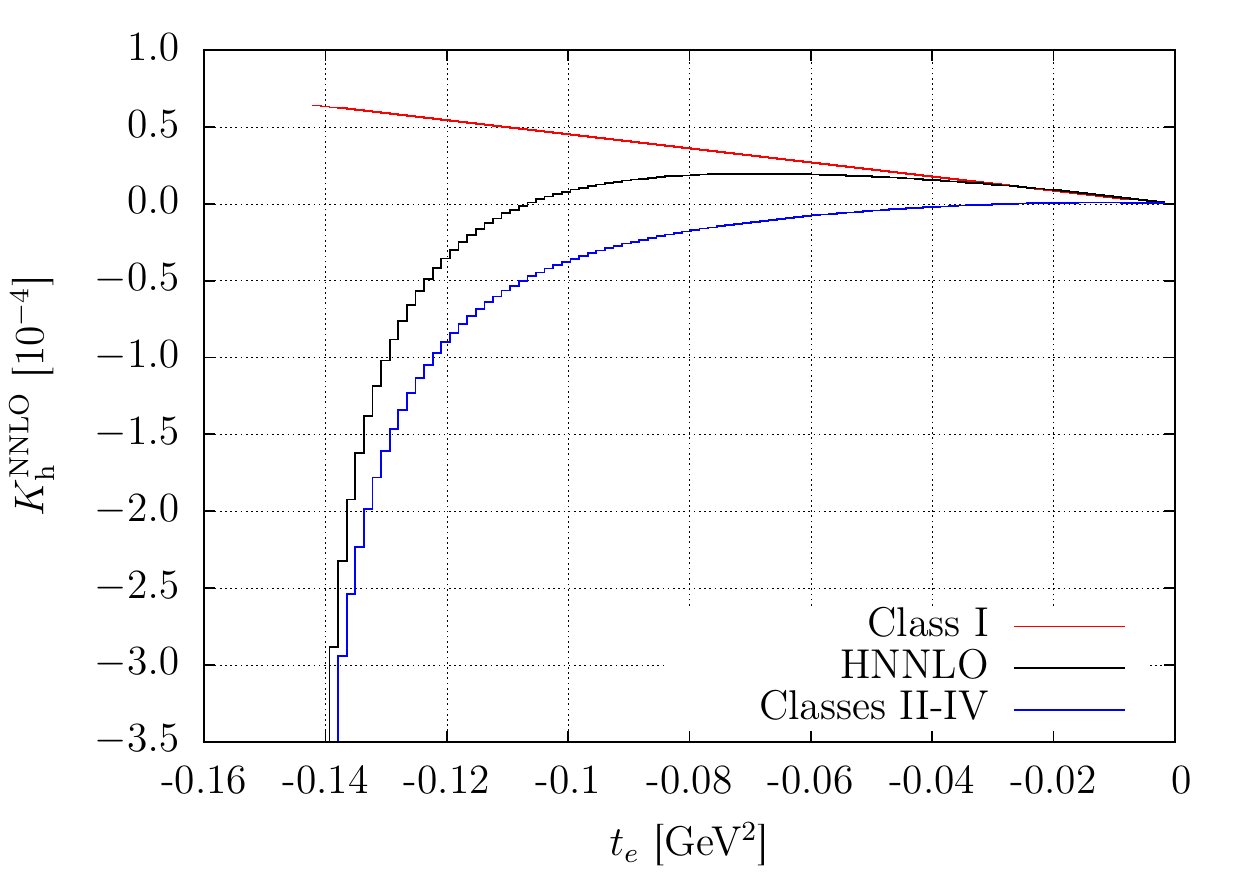}\\
    \includegraphics[width=1\columnwidth]{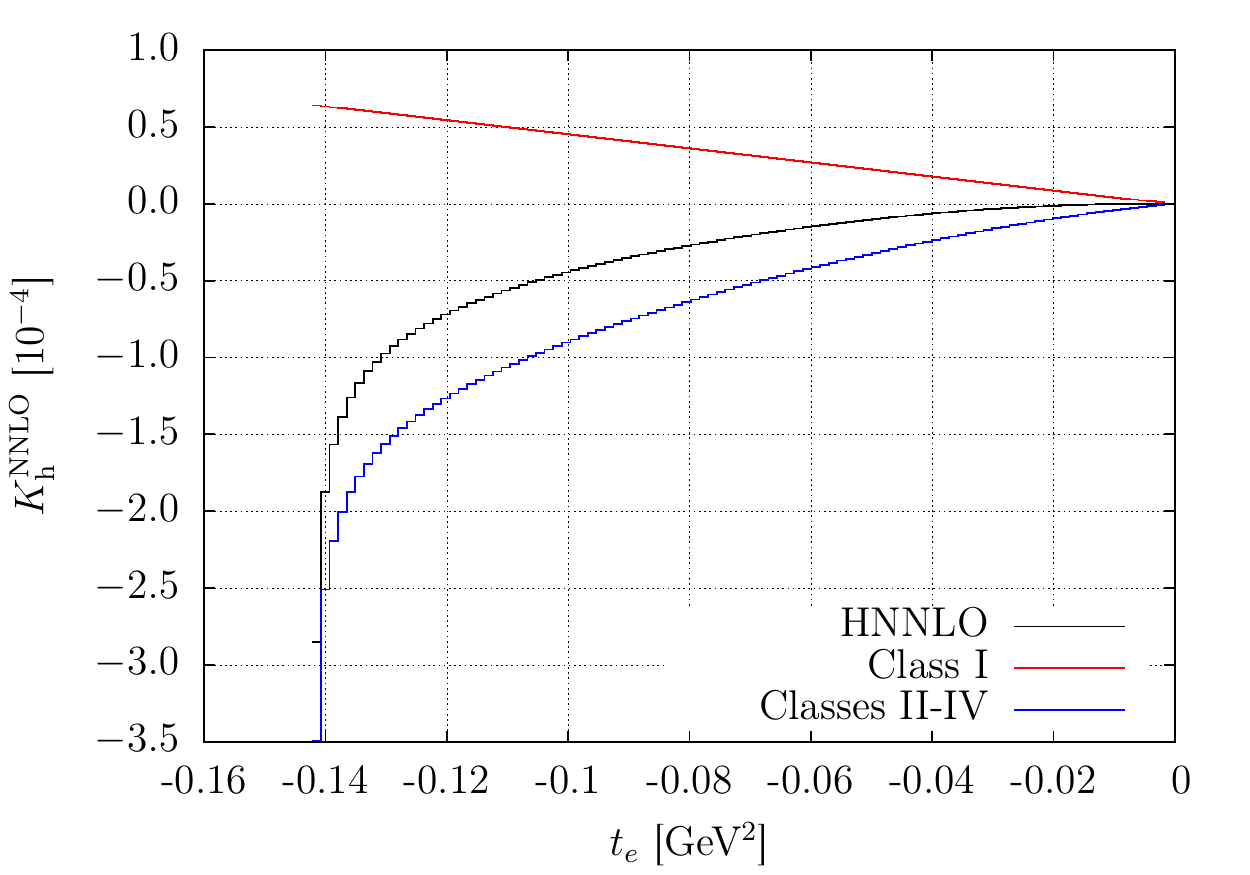}
  \caption{$K_\had^{\mathrm{NNLO}}(t_e)$ factor for a positive (upper panel) and negative (lower panel) muon beam of  energy $E_\mu =$ 150 GeV. The total hadronic NNLO correction are depicted in black, while the contributions of class I (II-IV) are shown separately in red (blue).}
  \label{fig:K}
\end{figure}

The ratio of the NNLO hadronic corrections to the $\mu e$ differential cross section, with respect to the squared momentum transfer $t_e$, and the LO prediction,
\begin{equation}
	K_\had^{\mathrm{NNLO}}(t_e) = \frac{d \sigma_\had^{\mathrm{NNLO}}}{dt_e} / \frac{d \sigma_0}{dt_e},
\label{eqn:RhoHNNLO}
\end{equation}
is shown in Fig.~\ref{fig:K} for the processes $\mu^+ e^- \to \mu^+ e^-$ (upper panel) and $\mu^- e^- \to \mu^- e^-$ (lower panel) for $E_{\mu} = 150$~GeV. 
The black lines indicate the total hadronic contribution arising from classes I--IV, while the blue ones show the sum of the contributions of classes II, III, and IV, but not I. The reason for this split is the following. The goal of the MUonE experiment is to determine $\Delta \alpha_\had(t) = -\Pi_\had(t)$, the leading hadronic contribution to the running of the effective fine-structure constant in the spacelike region, from $\mu e$ scattering data. In order to extract the NLO hadronic correction to the $\mu e$ differential cross section, given by~\eq{sigmaHNLO}, which contains $\Pi_\had(t)$, the experimental data will have to be subtracted, via a Monte Carlo event generator, of the total NNLO hadronic corrections (classes I--IV). If, instead of $\Delta \alpha_\had(t)$, one wants to extract the hadronic corrections to the resummed photon propagator, then the corrections of class I should not be subtracted from data, as their contribution to the differential cross section accounts for the second-order reducible hadronic contribution to the running of $\alpha(t)$.

The difference in $K^\mathrm{NNLO}_\had(t)$ between muon and antimuon is due to the box diagrams in classes II and IV, and to electron-muon interference terms in the real emission (class III). These contributions to the cross section are equal in size but with opposite sign for $\mu^+$ and $\mu^-$. The same pattern is observed at NLO~\cite{Alacevich:2018vez}.

Figure~\ref{fig:K} shows that, when the muon/antimuon beam has an energy of 150 GeV, for most of the kinematic region scanned by the squared momentum transfer $t_e$ the factor $K_\had^{\mathrm{NNLO}}(t_e)$ is of order $10^{-4}$--$10^{-5}$. These corrections are therefore larger than the $O(10^{-5})$ precision expected at the MUonE experiment. 
Moreover, our Fortran code, available upon request, can calculate the NNLO hadronic corrections to any $\mu e$ scattering differential distribution with arbitrary kinematical cuts and can therefore be implemented in future full NNLO $\mu e$ scattering Monte Carlo codes.

At NLO, the tiny contribution of the top quark to the vacuum polarization can be separated from the hadronic one. At NNLO, these contributions mix with each other. The plots in Fig.~\ref{fig:K} were obtained adding $\Pi_\mathrm{top}(q^2)$ to $\Pi_\had(q^2)$, so that the full top quark contribution has been included in the shown NNLO prediction. Its effect is however totally negligible.

As our calculation of the NNLO hadronic corrections to the $\mu e$ differential cross section is based on the hadronic $e^+ e^-$ annihilation data, the precision of our prediction is limited by the experimental error on the $R(s)$ ratio. We estimated the  uncertainty of our results induced by this error by comparing the values obtained with the libraries \texttt{alphaQEDc17} and \texttt{KNT18VP}. For each value of $t_e$, we found that the relative difference between the two calculations of $d\sigma_\had^\mathrm{NNLO}/dt_e$ is about 1\% or less. We therefore assigned to our $d\sigma_\had^\mathrm{NNLO}/dt_e$ predictions a relative uncertainty of 1\%, which corresponds to an error in $K^\mathrm{NNLO}_\had (t_e)$ of $O(10^{-6})$ or less, well below the precision expected at the MUonE experiment.

By employing the well-known one-loop analytic expression for $\Pi_\mathrm{l}(q^2)$ instead of $\Pi_\mathrm{h}(q^2)$, i.e.\ substituting the hadronic blob in Fig.~\ref{fig:classIV} with an electron or a muon loop, we compared our results for the vertices in class IV with the two-loop analytic expressions for the QED form factors of Ref.~\cite{Bonciani:2003ai}. We found perfect agreement between our numerical dispersive integrations and their explicit two-loop results (see also~\cite{Fael:2018dmz}).

\section{Conclusions}

In this letter we presented the NLO and NNLO hadronic corrections to the differential cross section for the processes $\mu^{\pm} e^- \to \mu^{\pm} e^- (+ \gamma)$, where (+$\gamma$) indicates the possible emission of photons.

We showed that, in a fixed-target experiment where the electron is initially at rest and the energy of the incoming muons or antimuons is 150 GeV, the corrections to the differential scattering cross sections with respect to $t_e$, the square of the difference of the initial and final electron momenta, are of order $10^{-4}$--$10^{-5}$ for most of the kinematic region spanned by
$t_e$. These corrections will therefore play a crucial role in the data analysis of future high-precision muon-electron scattering experiments like MUonE, whose goal is to reach a relative precision of order $10^{-5}$. The relative theoretical uncertainty of our predictions, induced by the experimental error of the hadronic $e^+ e^-$ annihilation data, is estimated to be about 1\% or less. It is therefore well below the precision expected at the MUonE experiment.

Making use of crossing relations, our results are also relevant to muon- and tau-pair production at present and future $e^+ e^-$ colliders operating at high and low energies.

~
\begin{acknowledgments}

{\em Acknowledgments} We are indebted to M.~Vitti for participating in the early stages of this work~\cite{MarcoVitti2018}. We would like to thank C.~M.~Carloni Calame, M.~Gravino, F.~Jegerlehner, A.~Keshavarzi, P.~Mastrolia, F.~Piccinini, A.~Primo, A.~Signer, W.~J.~Torres Bobadilla and G.~Venanzoni for useful discussions and correspondence.
We are also grateful to all our MUonE colleagues for our stimulating collaboration.
The work of M.~F.\ is supported by DFG through the Research Unit FOR 1873 ``Quark Flavour Physics and Effective Field Theories''. M.~P.\ acknowledges partial support by the MIUR-PRIN project 2010YJ2NYW and FP10 ITN Elusives (H2020-MSCA-ITN-2015-674896) and Invisibles-Plus (H2020-MSCA- RISE-2015-690575).
\end{acknowledgments}

\bibliographystyle{apsrev4-1}
\bibliography{BIB}
\end{document}